\documentclass[journal, onecolumn]{IEEEtran}
\pdfoutput=1
\usepackage{geometry}
\usepackage{amsmath}
\usepackage{graphicx}
\usepackage[hidelinks]{hyperref}
\usepackage[table,xcdraw]{xcolor}
\usepackage{rotating}
\usepackage{longtable}
\usepackage{multirow}
\usepackage{cite}
\usepackage{enumerate}
\begin{document}
\title{Socially-Aware  Evaluation Framework for Transportation}

\author{\IEEEauthorblockN{Anu Kuncheria\IEEEauthorrefmark{1}\IEEEauthorrefmark{2}\IEEEauthorrefmark{3},
Joan L. Walker Ph.D.\IEEEauthorrefmark{2}\IEEEauthorrefmark{3},
Jane Macfarlane Ph.D. \IEEEauthorrefmark{1}\IEEEauthorrefmark{3}}

\IEEEauthorblockA{\IEEEauthorrefmark{1} Smart Cities Research Center, Institute of Transportation Studies, University of California, Berkeley}

\IEEEauthorblockA{\IEEEauthorrefmark{2}Department of Civil and Environmental Engineering, University of California, Berkeley}

\IEEEauthorblockA{\IEEEauthorrefmark{3}Lawrence Berkeley National Laboratory \\
October 2021 \\
Corresponding email: anu\_kuncheria@berkeley.edu}}

\maketitle

\begin{abstract}
Technological advancements are rapidly changing traffic management in cities. Massive adoption of mobile devices and cloud-based applications have created new mechanisms for urban traffic control and management. Specifically, navigation applications have impacted cities in multiple ways by rerouting traffic on their streets. As different routing strategies distribute traffic differently across the city network, understanding these differences across multiple dimensions is highly relevant for policymakers. In this paper, we develop a holistic framework of indicators, called \textit{Socially- Aware  Evaluation Framework for Transportation} (SAEF), that will assist in understanding how traffic routing and the resultant traffic dynamics impact city metrics, with the intent of avoiding unintended consequences and adhering to city objectives. SAEF is a holistic decision framework formed as an assembled set of city performance indicators grounded in the literature. The selected indicators can be evaluated for cities of various sizes and at the urban scale. The SAEF framework is presented for four Bay Area  cities, for which we compare three different routing strategies. Our intent with this work is to provide an evaluation framework that enables reflection on the consequence of policies, traffic management strategies and network changes. With an ability to model out proposed traffic management strategies, the policymaker can consider the trade-offs and potential unintended consequences. 
\end{abstract}

\IEEEpeerreviewmaketitle

\section{Introduction}
Massive adoption of mobile devices and cloud-based applications have created new mechanisms for understanding how people move in urban environments. This new mobility data along with data generated by city infrastructure will provide cities with a more detailed view of traffic dynamics and allow them to play a more active role in managing urban performance. In addition, emerging connected and automated technologies have the promise to create more efficient solutions for city management. For example, the promise of autonomous and connected vehicle fleets in smart cities may provide the possibility of optimal traffic management through mass control of vehicle routes. These technologies will augment current mechanisms for traffic management in cities e.g. traffic lights, variable message signs, HOV lanes, and tolls. 

Traffic flows in urban environments are currently heavily influenced by real-time routing provided by various independent navigation systems (Google maps, Waze, Apple maps etc) \cite{hendawi_smart_2017 ,siuhi_opportunities_2016}, with up to 60\% of drivers using them daily \cite{noauthor_people_nodate, noauthor_popularity_nodate}. These systems add another level of control  that is not in coordination with existing infrastructure control. This has created undesired traffic dynamics, mostly driven by real-time route guidance systems, that often compromise safety and health in the neighborhoods affected \cite{gmt_your_nodate}.

Our goal is to develop a holistic framework of metrics that will assist in understanding how routing strategies and their resultant traffic dynamics impact city metrics, with the intent of avoiding unintended consequences and adhering to city objectives. Our framework, called \textit{Socially- Aware  Evaluation Framework for Transportation - SAEF}, is an evaluation framework with multiple measures related to urban performance, such as safety, equity and neighborhood congestion. The selected metrics can be evaluated for cities of various sizes and at urban scale.  The framework is designed to allow decision-makers to assess typical routing strategies and evaluate the potential impact on different aspects of a city. For illustrative purposes, we present the framework for four cities in the Bay Area. The routing strategies we evaluate are (1) user equilibrium in which travel time for each user is optimized, (2) system optimal travel time, and (3) system optimal fuel use. The impact of these optimization strategies on the Bay Area are generated using results from a mesoscopic simulation platform called Mobiliti \cite{chan_mobiliti_2018} that implements a Quasi Dynamic Traffic Assignment (QDTA) \cite{chan_quasi-dynamic_2021}. The QDTA algorithm partitions the day into 15 minute intervals, performs a static traffic assignment for each interval including trip accounting that allows for residual traffic from previous time interval. Although the framework was developed with the primary objective of evaluating the impact of traffic routing strategies, it may also be used as an evaluation tool for a wider array of transportation projects, such as infrastructure changes, connected traffic signals and traffic management projects. 
 
The remainder of the paper is organized as follows. In Section 2, a background of the existing transportation frameworks is provided. The design of our framework and indicators are presented in Section 3; the study methods, results and interpretation is elucidated in Section 4. Finally, the conclusions are discussed in Section 5, along with possible directions for future work.

\section{Literature Review }
\subsection{Traffic Routing in Cities}
There  has  been  much  previous  work  in  the  area  of  transportation modeling and traffic route choices \cite{doi:10.3141/1617-25,groot_toward_2015,ran_new_1993,zhu_linear_2015}. Understanding the distribution of traffic on local road network gained popularity in the past decade with route guidance systems rerouting traffic differently than before \cite{ackaah_exploring_2019,alaok_smart_nodate,paricio_urban_2019}. Most route guidance systems aim to provide an user with the least travel time route \cite{noauthor_predicting_nodate, mahmassani_network_1993}. This might mean taking people off the highway to local streets to save few minutes of travel time \cite{jou_route_2005}. Some services provide routes to users based on shortest fuel consumption. This could mean travelling at a consistent speed and thus prefer certain kinds of roads that maintains that speed limit \cite{alfaseeh_multi-factor_2020, ahn_network-wide_2013}. Studies evaluating the impacts of navigation systems model app based and non app based users differently to understand the resulting congestion patterns \cite{lazarus_decision_2018,cabannes_measuring_2018}. Most of them model a corridor or a small network with varying percentages of app users to show the increase in congestion as app users increase by using metrics like traffic flow, distance, and average marginal regret. Ahmed and Hesham modelled eco-routing as a feedback user equilibrium model for downtown LA and measured the outcome in terms of fuel consumption and congestion levels \cite{elbery_city-wide_2019}. It could be seen from the previous studies that different strategies has different impacts across the city network in terms of time and distance. However no previous studies have explored the impacts of routing holistically on multiple city dimensions. 

\subsection{Evaluation Frameworks in Transportation}
We conducted an extensive review of transportation literature to identify frameworks or indicators that can be used to assess the impacts of traffic routing strategies on cities. Due to lack of specific frameworks designed for routing, we reviewed the general frameworks in the transportation domain and assessed its usability for routing evaluation. 

European Commission’s CITYkeys framework developed a set of city performance measures at city level and project level \cite{ahvenniemi_what_2017}. The framework was focused on five major themes of people, planet, prosperity, governance, and propagation. Out of the 116 key performance indicators, the ones concerning transportation were in car waiting time, reduction in traffic accidents, quality of public transportation, improved access to vehicle sharing solution, extended bike route network, reduced exposure to noise pollution, reduction in annual energy consumption. HASTA framework measures sustainability for a transportation project based on three dimensions of economic, environmental and social and six sustainability indicator groups for Swedish cities \cite{koglin_measuring_2011}. They came up with a total of 83 indicators. Additionally, considerable work is ongoing through the International Standards Organization (ISO), European Committee for Standardization, and BSI to establish proper standards in smart urban development project evaluation. ISO 37122:2019 provides a framework for resilient city with 19 themes and multiple indicators for each theme. The transportation theme has 14 indicators, mostly focusing on real-time technology, electric fleet, and integrated payment systems (ISO 37122:2019). 
In addition to the frameworks, there is a vast literature on transportation indicators used for evaluation of new projects.  Specific to smart cities and transportation, Orlowski and Romanowska developed a set of smart mobility indicators that encompass the following domains: technical infrastructure, information infrastructure, mobility methods, and vehicles used for this purpose and legislation with 108 metrics \cite{orlowski_smart_2019}. Benevolo et al. (2016) investigated the role of ICT in supporting smart mobility actions, influencing their impact on the citizens' quality of life, and on the public value created for the city as a whole \cite{benevolo_smart_2016}. They provided an action taxonomy considering three aspects of smart mobility actors including the main agents moving the smart initiatives, the use and intensity of ICT and the benefits of smart mobility actions on smart goals , to investigate the role of ICT  in supporting smart mobility actions, influencing their impact on the citizens’ quality of life and on the public value created for the city as a whole. 

The existing frameworks cannot be directly used for routing evaluation as they are general, considering the entirety of transportation or entirety of smart cities. Only few works have reasoned on more complex aspects, such as how different indicators interact reciprocally, what benefits they generate, how they affect citizens' quality of life, what problems they solve and what is created. Further, they do not capture factors specific to the impact of routing strategies on cities. While the aspects of safety and neighborhood are discussed in some metrics, it is mostly qualitative and subjective in nature. Key challenges are centered on selecting suitable evaluation methodologies to evidence urban value and outcomes, addressing city’s objectives. With this focus, our framework grounded on literature develops a set of themes and indicators for cities to evaluate impacts of routing. To the best of our knowledge, this is the first time a holistic evaluation using multiple themes is used to identify the impacts of different routing strategies for a large scale network with cross city comparisons. 

\section{Socially- Aware  Evaluation Framework for Transportation}
\subsection{Framework Design}
\begin{figure}[ht]
    \centering
    \includegraphics[width=0.9\textwidth]{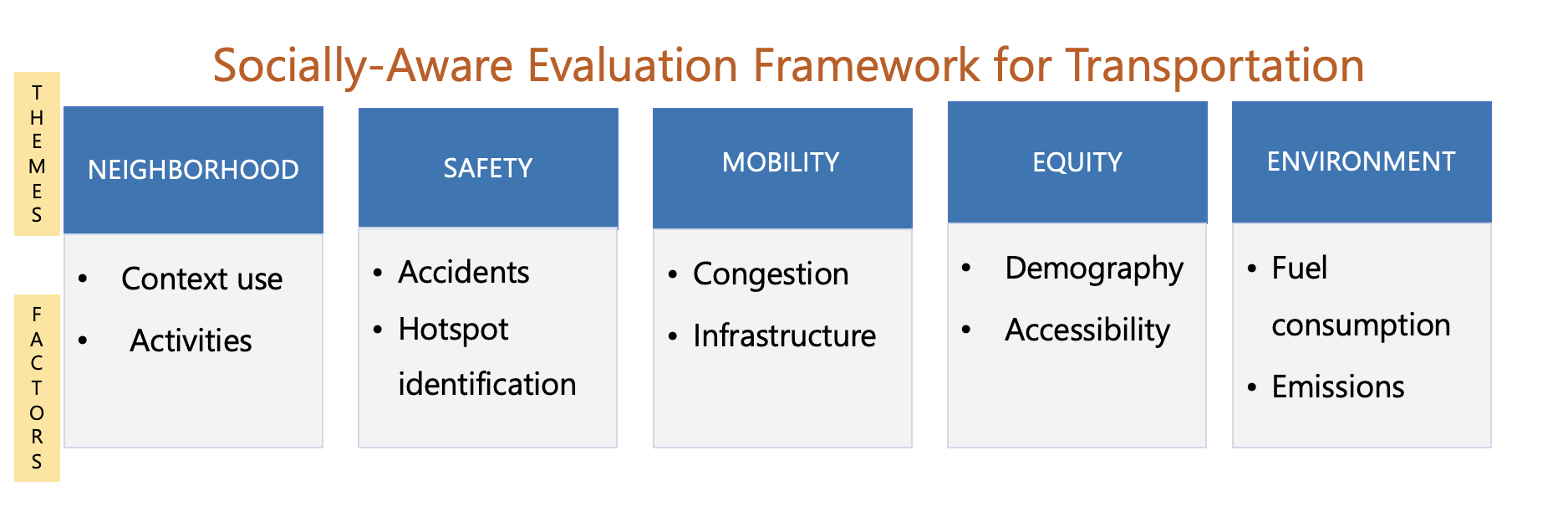}
    \caption{Themes in the framework}
    \label{fig:Themes}
\end{figure}

\begin{figure}[ht]
    \centering
    \includegraphics[width=1\textwidth]{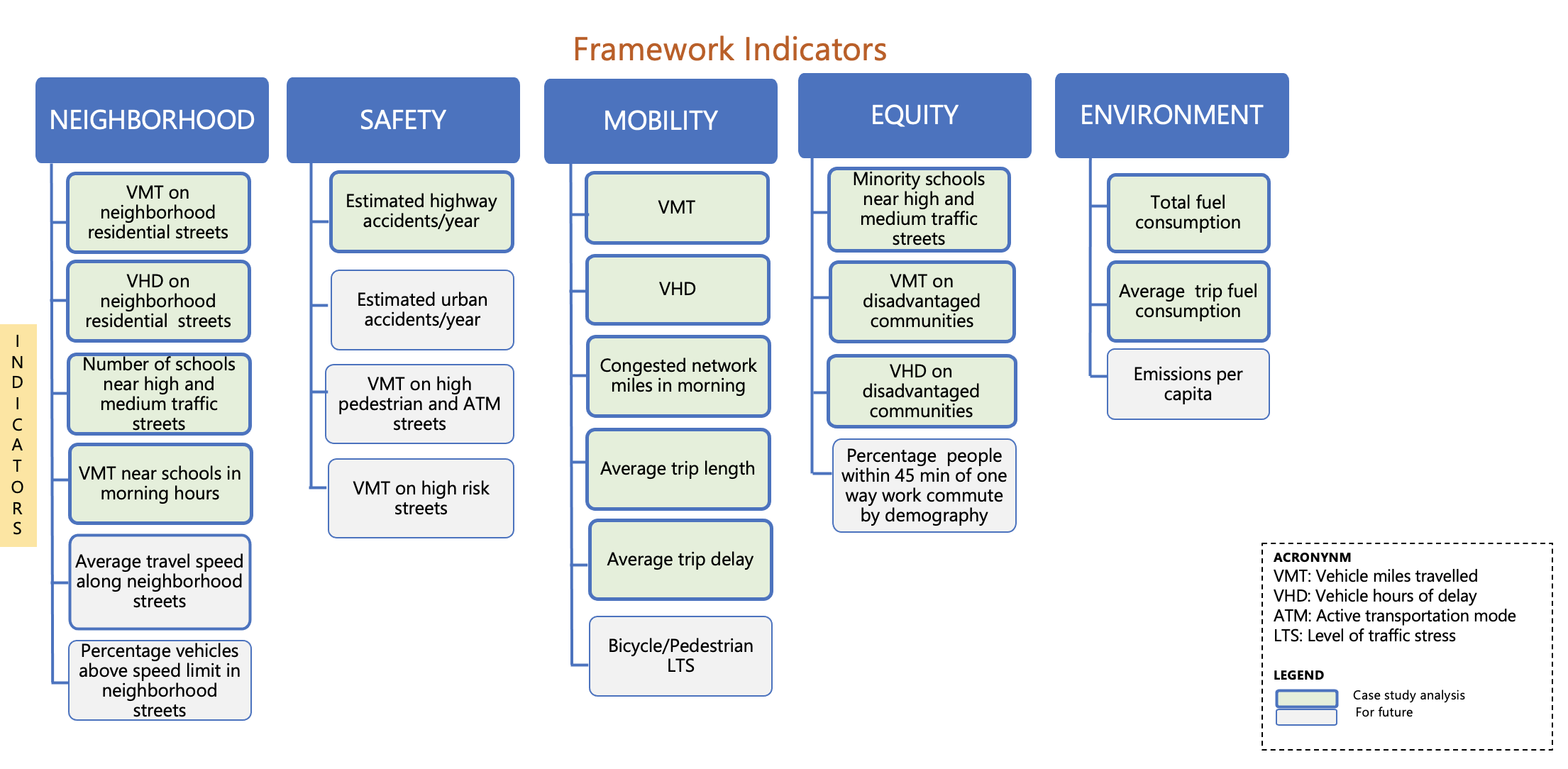}
    \caption{Indicators for each theme}
    \label{fig:indicators}
\end{figure}

To explore the city level impacts of various routing strategies, we develop a holistic framework called \textbf{Socially- Aware  Evaluation Framework for Transportation - SAEF}. We define the term 'socially-aware' to include five complementary themes of neighborhood, safety, mobility, equity and environment (Figure \ref{fig:Themes}).The themes are assembled from a set of city performance indicators grounded in literature \cite{onatere_performance_2014,TOTHSZABO20122035, koglin_measuring_2011}. Each theme identifies factors that are likely key concerns for that theme. For example, accidents are a concern when considering safety, similarly emissions are a concern when considering environmental quality. Based on these factors, we defined indicators/metrics  (Figure \ref{fig:indicators}) that could be derived from real world or simulated data. These indicators were culled from a thorough review of literature from domestic and international, city policies and international organizations related to transportation. A list of over 100 indicators with wide range of scale and use were identified and a manageable set of indicators that we assessed as being important as well as measurable were created for each theme.  This resulted in 23 indicators: 6 were selected as reflections of neighborhood quality, 4 for safety, 6 for mobility, 4 for equity, and 3 for environment quality. 
This framework can then provide a basis for:
\begin{itemize}
    \item comparing indicators as a function of specific traffic management strategies, 
    \item attaching weights to the metrics that reflect planning objectives,
    \item stressing the relationship between policy goals and intended city level impact, and 
    \item monitoring progress towards long term policy goals.
\end{itemize}


\subsection{Operationalizing the Framework}
Operationalizing the framework is the key to its usability for cities managers. SAEF provides a set of measurable indicators in context of a broad framework. It provides a mechanism for prioritizing city’s objectives with an understanding of the trade offs that must be made to achieve certain objectives, for example if a traffic management strategy attempts to reduce fuel consumption it may result in more vehicles travelling through neighborhoods. City planners can then determine whether the reduced emissions are higher priority than the likely safety costs of higher traffic flows on neighborhoods streets. They can choose specific themes from the framework based on their own objectives and values and evaluate them in context of proposed traffic management strategies.

\begin{figure}[ht!]
    \centering
    \includegraphics[width=0.9\textwidth]{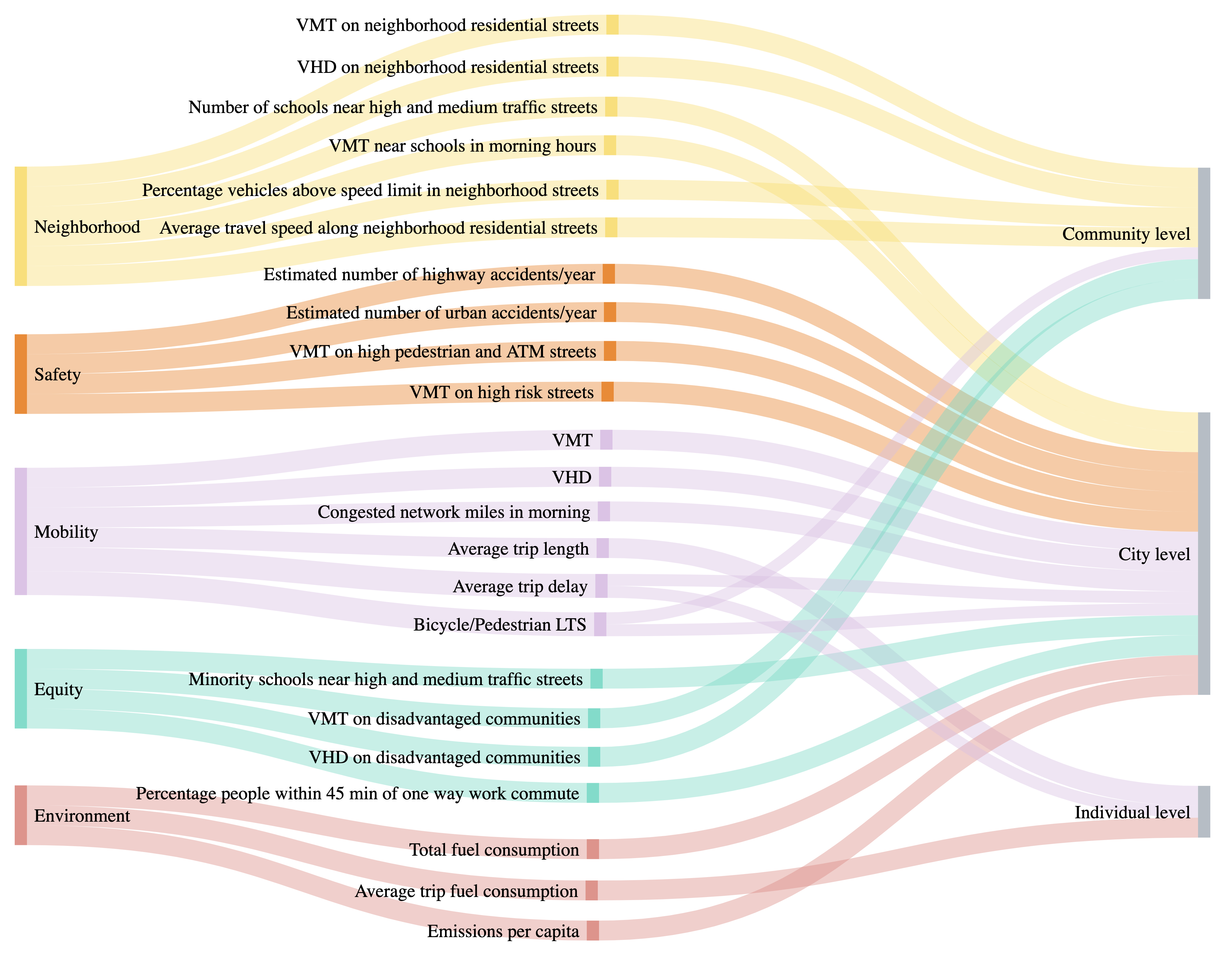}
    \caption{SAEF Indicators and their spatial levels. }
    \label{fig:spatiallevel}
\end{figure}

Each indicator is detailed in Appendix \ref{lab:appendixA}, including a description, the unit of measurement and spatial/temporal levels. The indicators are structured in spatial and temporal dimensions. The spatial dimension is partitioned into individual, community, and city levels (Figure \ref{fig:spatiallevel}). There are 2 temporal dimensions -  peak hours or entire day -  based on the indicator relevance. The list of indicators is not intended to be exhaustive. It can be revised as new information becomes accessible.

The next section addresses how to quantify the indicators. Once the indicators are quantified, the decision makers can weight them to generate an aggregate score for decision making. The weighting should reflect the goals of the city and rationalize the values of the indicators. Visualization of the indicators in context of various strategies is an alternative method. 

\section{Application of the Framework}
\subsection{Traffic Simulation}
\begin{figure}[ht!]
    \centering
    \includegraphics[width=0.8\textwidth]{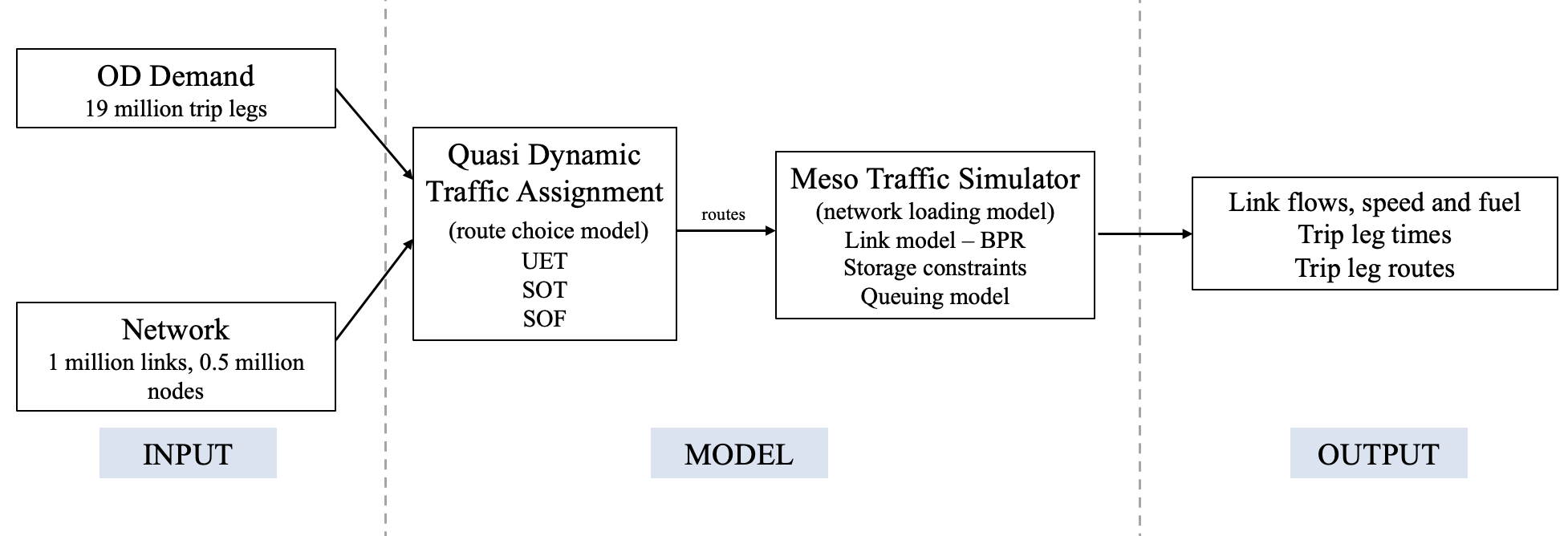}
    \caption{Mobiliti simulation modelling}
    \label{fig:mobiliti_framework}
\end{figure}

In order to evaluate impacts of various routing strategies, we use the results from an agent based simulator Mobiliti \cite{chan_mobiliti_2018}. Mobiliti is an urban-scale simulation platform for evaluating network traffic dynamics. Mobiliti provides three optimization methods based on standard traffic assignment algorithms. Specifically, it implements a Quasi-Dynamic Traffic Assignment (QDTA) \cite{chan_quasi-dynamic_2021}. The main components of the simulator are detailed in Figure \ref{fig:mobiliti_framework}. For a given demand, route choice is generated in the QDTA step based on the principles of user equilibrium or system optimal. The optimization objectives are travel time or fuel (refer Appendix \ref{lab:appendixB} for the specific objective functions). The network loading is generated using the Mobiliti network simulation, with trips paths assigned by the optimization step. The transportation network for Bay area has over 1 million links and 0.5 million nodes. The trip demand is defined by a travel demand model which for the Bay Area is the SFCTA CHAMP 6 model accounting for $\sim$19 million trips during a 24-hour period \cite{noauthor_sf-champ_nodate}. 

\begin{figure}[ht!]
    \centering
    \includegraphics[width=0.7\textwidth]{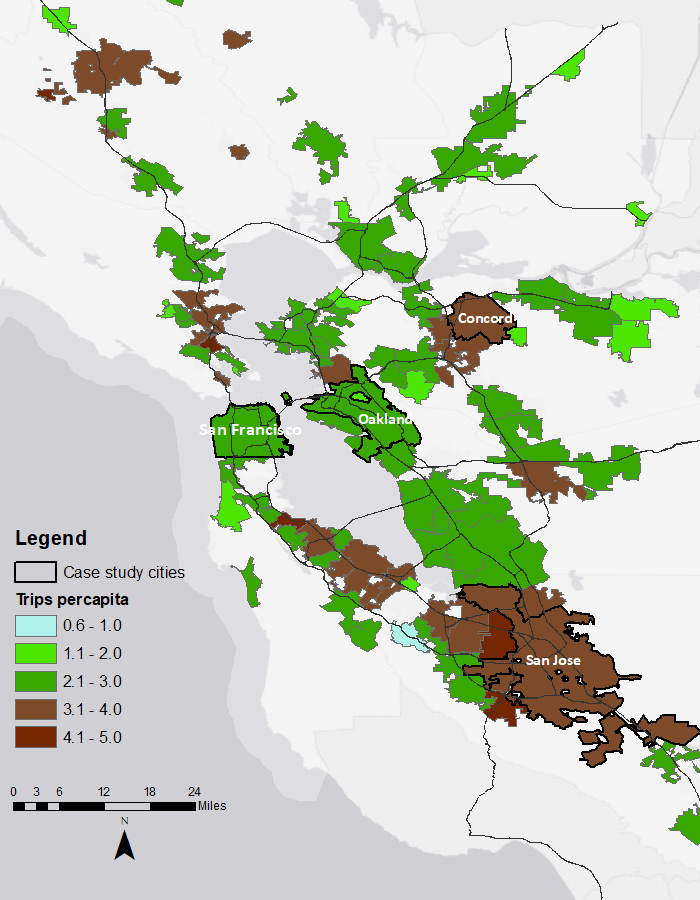}
    \caption{Bay area cities colored by percapita trips.}
    \label{fig:citytrips}
\end{figure}

We studied three optimization scenarios: (1) user equilibrium travel time (UET), (2) system optimal travel time (SOT), and (3) system optimal fuel (SOF). The typical optimization for trip level shortest travel time (UET) is  compared against the system optimal strategies. The embedded complexity of transportation networks naturally requires trade-offs when optimizing for different objectives. Our goal with the framework is to provide context for understanding the trade-offs. The results are examined for four cities: San Jose, San Francisco, Oakland and Concord. These cities were selected to represent different-sized conurbations, relevant to their population and geographical area. Figure \ref{fig:citytrips} shows all Bay area cities with the chosen four case study cities highlighted. 

\begin{table}[ht!]
\centering
\caption{Case Study Cities}
\label{tab:casecities}     
\begin{tabular}{|l|r|r|r|r|}
\hline
\noalign{\smallskip}
City  & Area  & Population & Population density & Trips per capita \\ 
& (sq.mile) &  & (person/sq.mile) & 
\\
\noalign{\smallskip}
\hline
\noalign{\smallskip}
San Jose & 185 & 1,021,795 & 5,524 & 3.1\\
San Francisco  & 49 & 874,961 & 17,847 & 2.2\\ 
Oakland & 58 & 425,097 & 7,333 & 2.7 \\
Concord & 31 & 129,183 & 4,105 & 3.5 \\
\noalign{\smallskip}
\hline
\end{tabular}
\end{table} 

\subsection{Network Typologies}
\label{sec:networktypologies}
Flow distribution on roads will differ based on how the system is optimized. Hence understanding the distribution of traffic flows on road types is relevant for impact elevation. Traditionally, road classification systems are based on mobility and access for vehicular use \cite{road_class}. This study seeks to incorporate the local context of streets, which includes how the dynamics of the vehicular traffic impacts localized populations. To this end, we created a road classification scheme based on the principles of USDOT complete street guidelines \cite{noauthor_complete_nodate}. Complete streets provides guidelines for the design and operation of streets to enable safe use and support mobility of all citizens. Adoption of these guidelines is underway by a variety of cities. For example, San Francisco has classified their street based on land use characteristics, transportation roles and special characteristics (eg. alleys, parkways). This resulted in 16 classes that were designed using extensive community surveys and manual labelling \cite{noauthor_better_nodate}. Similarly, San Jose classified streets into 8 types based on a street’s primary function and adjacent land use context \cite{noauthor_sj_2018}. Because we wish to compare our indicators across all of Bay Area cities, we created an alternative classification scheme based on parcel level zoning data and street functional classes. This classification scheme, of 8 types, allowed us to then partition and develop improved metrics that will be helpful for evaluating the themes of our framework. 

\begin{figure}[ht!]
    \centering
    \includegraphics[width=0.7\textwidth]{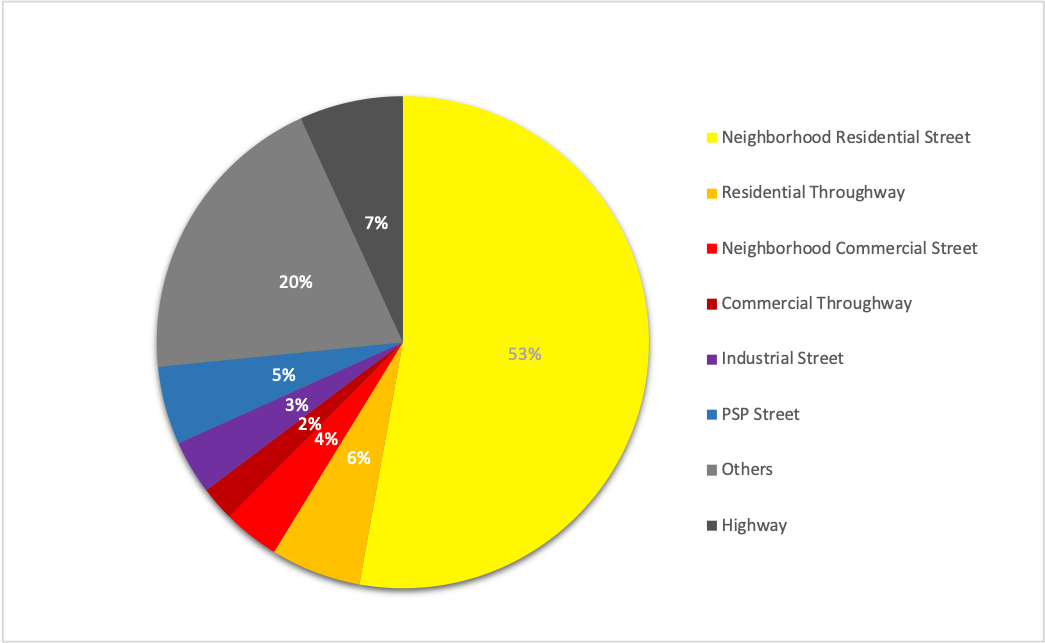}
    \caption{Network typologies for Bay Area. Neighborhood residential streets constitutes the highest share followed by highways. Individual cities reflect similar partitioning.}
    \label{fig:bayareatypes}
\end{figure}

\begin{figure}[ht!]
    \centering
    \includegraphics[width=\textwidth]{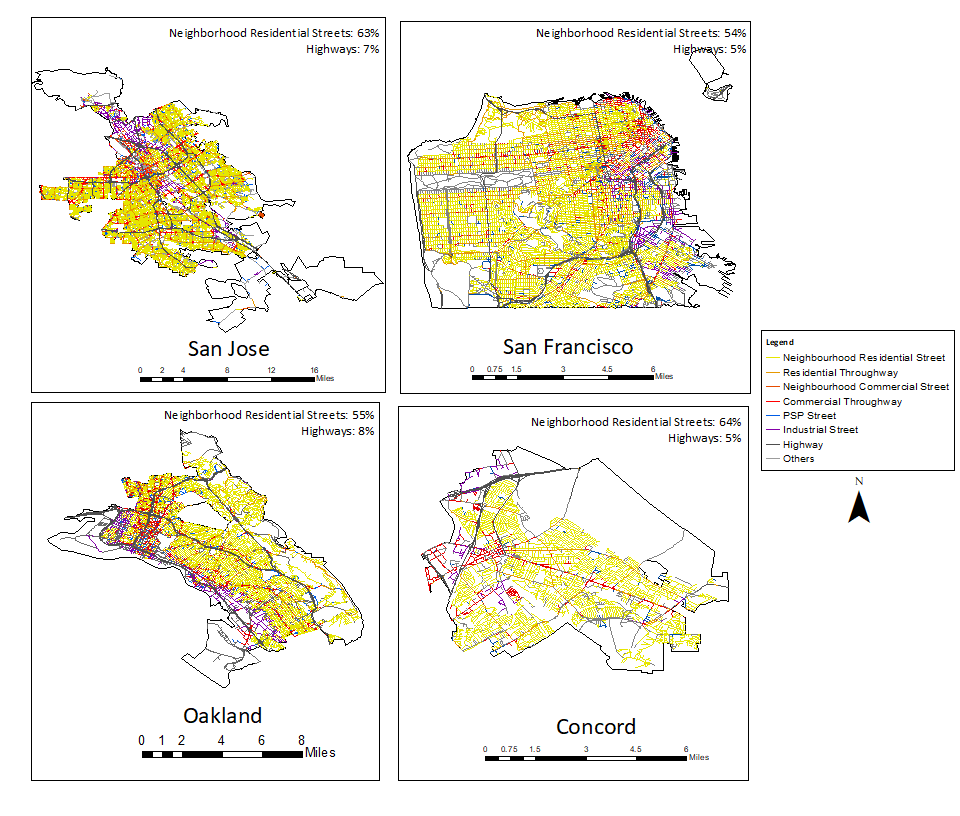}
    \caption{Network Typologies for Oakland, San Jose, San Francisco and Concord. Oakland has the highest percentage of highways and Concord has the highest percentage of neighborhood residential streets.}
    \label{fig:cities_links}
\end{figure}

Our classification uses the Mobiliti road network, which was derived from a professional map from HERE Technologies \cite{noauthor_here_nodate}. The HERE technologies map includes definitions of five functional class roads defined in Appendix \ref{lab:appendixC} Table \ref{tab:fcclass}. For identifying the transport context, links are classified into three types - highways, throughways, and neighborhood streets. These are based on the functional classification and speeds: highways group higher functional class links of 1, 2, and 3 with speeds greater than 50 mph, throughways group rest of class 3 and 4 links that carry greater volumes and higher speeds of vehicle traffic, and neighborhood streets group class 5 links with lower speeds and volumes. For identifying land use context, parcel level zoning data obtained from Metropolitan Transportation Commission (MTC) is used. There are 1,956,207 parcels in the analysis region grouped into either of the five land uses - residential, commercial, industrial, public-semi public and others. For each link, the side use is determined based on the zoning of the adjacent parcels. If the link is associated with more than one parcel, the use of the largest parcel is assigned to the link. Based on the transport and land use context for each link, 8 street types are established. This classification composition for Bay Area streets is shown in Figure~\ref{fig:bayareatypes}. Appendix \ref{lab:appendixC} Table \ref{tab:link_class} details the network lengths associated with the classification. The classification, while slightly coarser than a more detailed partitioning, does not involve the complexity of surveys and manual labeling and can still provide a good understanding of the network typologies. Figure \ref{fig:cities_links} shows the resulting context based link classification for the four cities of interest.

\subsection{Methods for Quantifying Indicators}
Quantifying the indicators described in the framework can be accomplished by using existing models, developing new models, and accessing city data, eg. zoning use, numbers of accidents, traffic flow and speeds from highway and city detectors, locations of schools. For our initial approach, we chose to confine our attention to the development of the broad framework as discussed in the previous section and examine a subset of indicators. A list of indicators evaluated for our case study cities and their corresponding method for evaluation is provided in Table \ref{tab:summary_indicators}. The Mobiliti simulation results for each optimization scenario described in the previous section provided link level flow and speeds for our evaluations.
\begin{table}[ht!]
\centering
\footnotesize
\caption{Methods used to evaluate indicators}
\label{tab:summary_indicators}     
\begin{tabular}{|p{2.5cm}|p{6.5cm}|p{6cm}|}
\hline
\noalign{\smallskip}
Theme & Indicator & Method  \\ 
\noalign{\smallskip}
\hline
\noalign{\smallskip}
Neighborhood & VMT on neighborhood residential streets, VHD on neighborhood residential streets & Neighborhood residential streets are identified as described in section \ref{sec:networktypologies}\\

 & Number of schools near high and medium traffic streets, VMT near schools in morning hours & Section \ref{sec:neighborhoodindicators} \\

Safety & Estimated number of highway accidents/year & Section \ref{sec:safetyindicators} \\

Mobility & VMT, VHD, Congested network miles in morning, Average trip length, Average trip delay & Section \ref{sec:mobilityindicators}\\

Equity & Minority schools near high and medium traffic streets, VMT on disadvantaged communities, VHD on disadvantaged communities & Section \ref{sec:equityindicators}\\

Environment & Total fuel consumption, Average trip fuel consumption & Section \ref{sec:mobilityindicators}\\
\noalign{\smallskip}
\hline
\end{tabular}
\end{table} 

\subsubsection{Neighborhood Indicators}
\label{sec:neighborhoodindicators}
Impact near schools is an important indicator in the neighborhood theme. Exposure to traffic-related air pollutants has been associated with a range of adverse long term and short term health effects. Long term traffic related air pollution can cause breathing and mental health problems, and in the short term, increased traffic flow near schools pose accident risk and congestion around schools. Previous studies have demonstrated that it is preferable to locate schools in areas with higher percentages of local roads in order to reduce the exposure to air pollutants \cite{yu_planning_2015}. However, vehicle route optimizations may have diverted traffic to local roads and inadvertently change the predicted exposure levels. We examine these impacts by a) identifying high and medium traffic flow links around schools, and b) estimating the increased vehicular traffic during morning schools hours. We further examine whether minority schools absorb more of this exposure than their counterparts.  

School data on the location and characteristics of 1849 public schools in the Bay Area, grades pre-kindergarten through 12 was obtained from  Elementary/Secondary Information System (ElSi), which is a web application of National Centre for Education Statistics (NCES) using data from the Common Core of Data (CCD) for the year 2018-2019.  We evaluated roads within a 250m radius of each school \cite{kingsley_proximity_2014}. Studies suggest that Average Daily Traffic (ADT) greater than 50,000 creates high exposure to traffic emissions, and between 25,000 to 50,000 is considered medium exposure \cite{green_proximity_2004, wu_proximity_2006}. For all links in the buffer of each school, ADT counts were calculated to determine if these thresholds were exceeded. In addition, we identify the effects of increased traffic flow during the morning 7-8 am hours when children are likely to be present, by calculating the vehicle miles travelled (VMT) for the same links in the 250m buffer. We then stratified by socio-economic indicators of the school. A school is defined as a minority school if the majority (75\% of higher) of students in the school are minority (Black, Hispanic, Asian/Pacific Islander, or American Indian/Alaska Native).

\subsubsection{Safety Indicators}
\label{sec:safetyindicators}
Estimating the accidents for highways and local roads is based on a number of factors like traffic flow, road geometry,and road type \cite{noauthor_road_nodate}. For the state of California, Caltrans has developed safety performance functions (SPF) to estimate the occurrence of accidents on highways \cite{shankar_methods_2015}. The safety performance function calculates the estimated number of accidents per year on a road segment as a function of  average daily traffic (ADT) and length of the road. We use the Type 1 SPF which is specified as:
\begin{equation}
\lambda_i = \alpha + ln(length)_i + \beta ln(ADT)_i 
\end{equation}
where: $\lambda$ is the estimated number of accidents per year, $ADT$ is the Average daily traffic. The model parameters estimated using Pems data is described in Appendix \ref{lab:appendixD}. 

\subsubsection{Mobility and Environment Indicators}
\label{sec:mobilityindicators}
Mobility indicators used in the framework include key system performance metrics eg. vehicle miles travelled (VMT), vehicle hours of delay (VHD), congested urban VMT, as well as, trip level metrics, eg. average trip distance and time.  VMT is typically a system performance measure reported for overall analysis. It is calculated as the vehicle flow on a link multiplied by the link lengths. VHD is the delay per vehicle for a given segment, multiplied by the total number of vehicles, where delay per vehicle is calculated as actual travel time minus the free flow travel time. Congested network miles is calculated as the length of network congested, where congestion is defined as volume over capacity greater than or equal to 1 for a link. For trip level metrics, the average travel distance and average time is calculated from Mobiliti trip data. For the environment theme, total system fuel consumption and average trip fuel consumption are also calculated by Mobiliti. 

\subsubsection{Equity Indicators}
\label{sec:equityindicators}
\begin{figure}[ht!]
    \centering
    \includegraphics[width=0.85\textwidth]{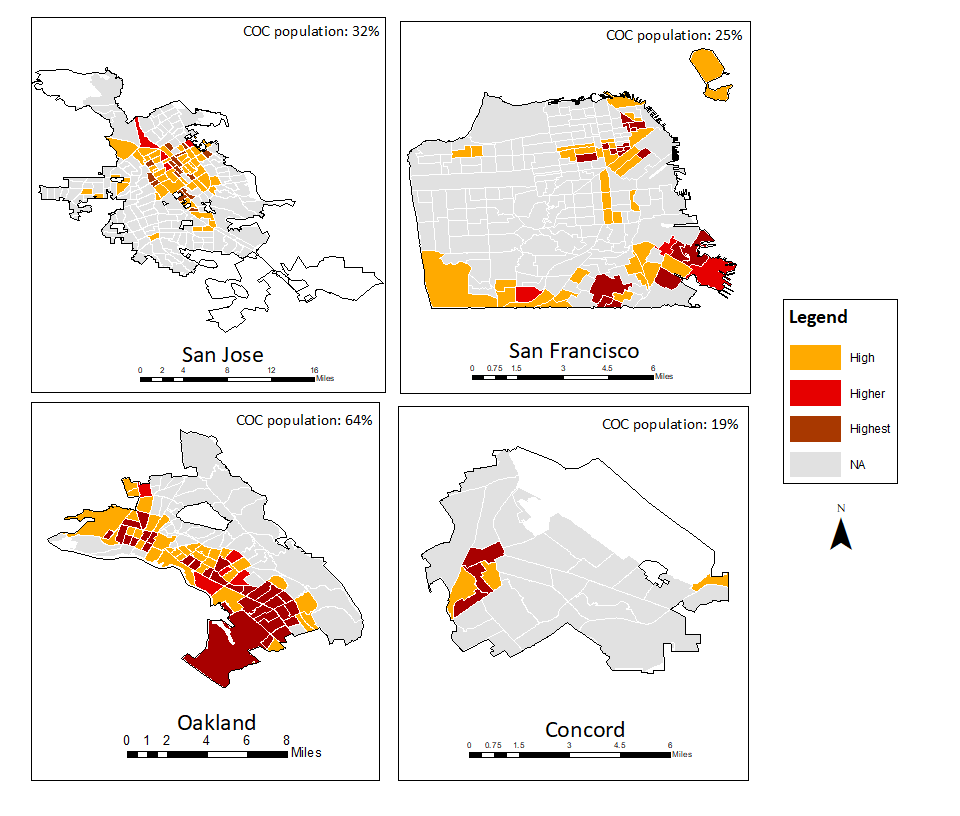}
    \caption{Theme Equity: Communities of Concern census tracts in the case study cities. Oakland has the highest percentage of population living in these census tracts and Concord the least.}
    \label{fig:coc}
\end{figure}

Transportation equity is an important consideration when evaluating routing strategies. Disadvantaged communities often bear disproportionate impacts of traffic exposure. To evaluate our equity measures, we identify disadvantaged communities from MTC's communities of concern classification. The 2020 update of MTC's Communities of Concern (COC) is based on on 2014 - 2018 American Community Survey (ACS) 5-year tract level data. The MTC COC definition is based on 8 disadvantage factors and respective thresholds \cite{noauthor_mtc_nodate}. The factors are minority, low income, limited English proficiency, zero vehicle households, senior citizens, people with disability, single parent family, and severely rent-burdened households. All census tracts with concern class categorised as high, higher and highest are considered as part of our disadvantaged group (Figure \ref{fig:coc}). We calculate the proportion of VMT and VHD on the links in COC census tracts and compare with the proportion of population in those census tracts. ACS 5-year estimates for 2014 - 2018 is obtained for census tract population \cite{bureau_american_nodate}.

\subsection{Evaluation Results}
\begin{figure}[ht!]
    \centering
    \includegraphics[width=1\textwidth]{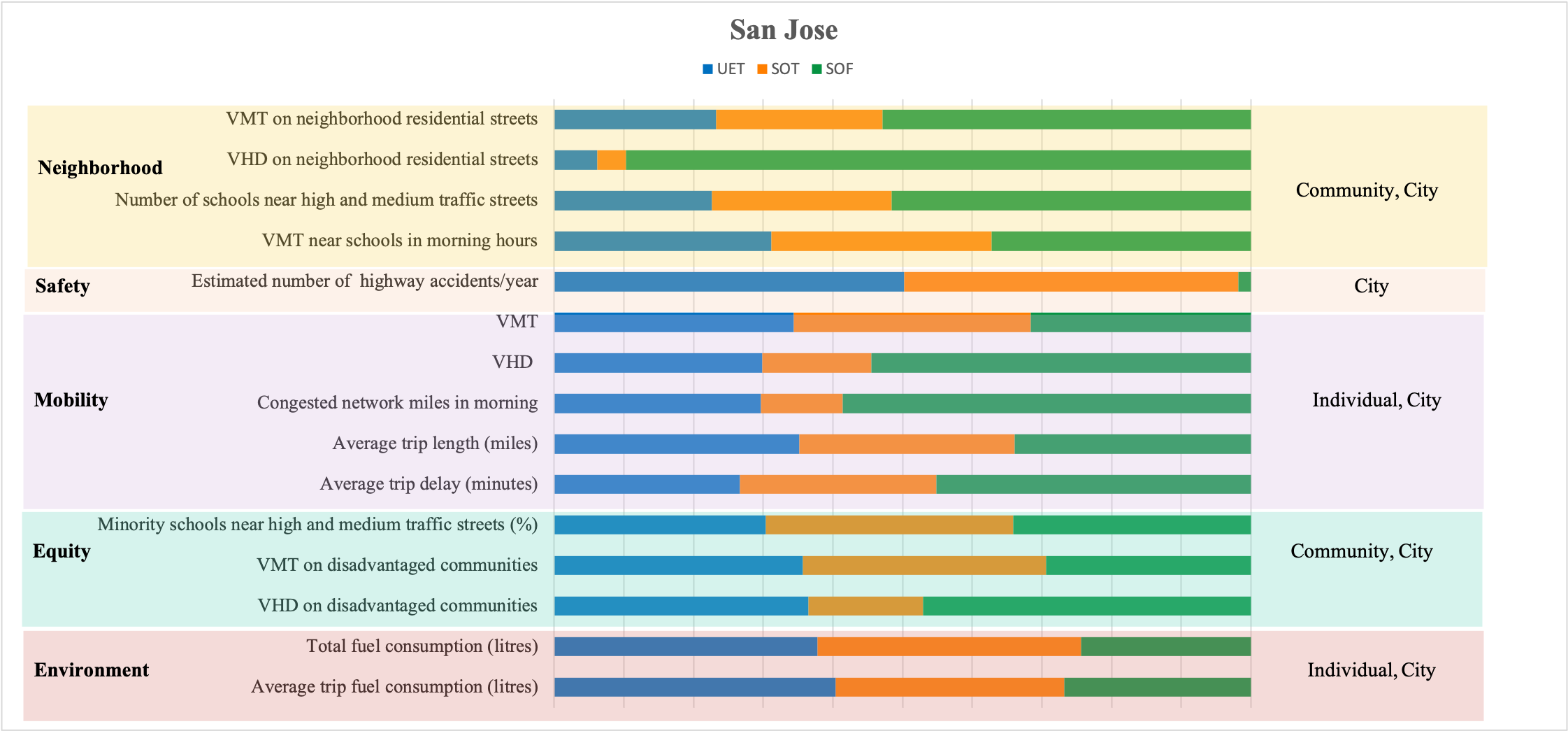}
    \caption{SAEF Indicators for San Jose. Blue, orange and green represents the UET, SOT and SOF indicator values.}
    \label{fig:metric_comparisons}
\end{figure}

For each city, fifteen indicators from the SAEF framework are calculated and compiled using a stacked bar chart for visual comparison across the routing optimization scenarios. Figure \ref{fig:metric_comparisons} presents the results for San Jose. Summary table and charts for the three other cities are provided in the Appendix \ref{lab:appendixE}. For added insights, we compare optimization scenarios for each of the indicators, similarly we also compare across our four focus cities. For the rest of this section we will discuss results for San Jose. For comparison of optimizations, we take UET optimized routes as our baseline.

\subsubsection{City Level Indicator Comparisons}
\noindent Theme: Neighborhood
\begin{enumerate}[a.]
    \item VMT disproportionately increases on neighborhood residential streets with SOT and SOF derived routes. While the total system VMT reduces with SOT and SOF, the residential VMT increases for both cases. Additionally the share of VMT carried by these streets (out of the total) also increases with SOT and SOF. Neighborhood residential streets account for 4\% of the total VMT in our baseline UET and increases to 5\% and 11\%  with SOT and SOF respectively.

    \item Neighborhood residential streets VHD decreases with SOT and increases with SOF. However the share of delay carried by these streets out of the total system delay increases in both SOT and SOF compared to baseline showing both optimizations has attained its goal by routing more vehicles through previously less used streets. In baseline, these streets carry 62 hours of delay as opposed to 41 in SOT and 900 in SOF. Looking at the percentage share of these delays with respect to total, in baseline it is 0.58\% of the total city VHD, which slightly increases with SOT (0.73\%) and more than doubles with SOF (4.5\%). This increase in system delay with SOF reflects highway delay being decreased significantly. 

    \item The number of schools exposed to high and medium traffic increases significantly with SOF due to traffic shift to local roads. Exposure to high and medium traffic occurs for 9\% of schools in our baseline UET. This percentage slightly increases with SOT; and doubles with SOF. VMT in the 250m buffer zone of school also increases with SOT and SOF as compared to baseline. 
\end{enumerate}

\noindent Theme: Safety
\begin{enumerate}[a.]
    \item Estimated number of highway accidents significantly reduces with SOF, while it remains similar for SOT and baseline. The obvious reason for the large reduction is the shift of highway traffic to local roads with the SOF optimization. It will be important to include an estimate of urban accidents, which we aim to add for future analysis. 
\end{enumerate}

\begin{figure}[ht!]
    \centering
    \includegraphics[width=1\textwidth]{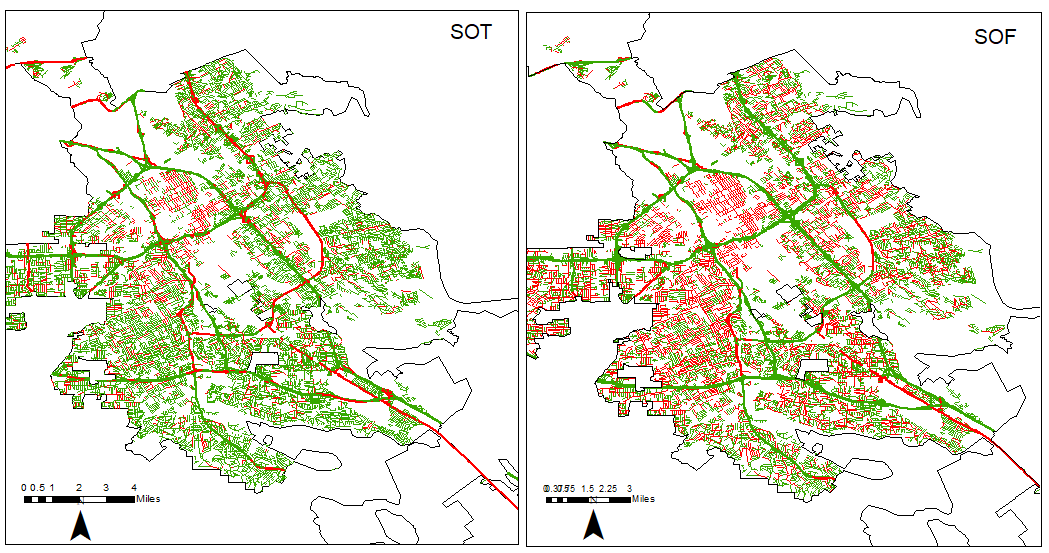}
    \caption{Theme Mobility: The figure shows the difference in VMT for SOT (left) and SOF (right) compared to baseline UET for San Jose. The red and green represents increase and decrease in VMT for each link respectively. The thickness of the links represents the magnitude of difference. Only highways and neighborhood residential streets are shown. It can be observed from the map that SOF has shifted traffic from highways to residential streets as indicated by red.}
    \label{fig:vmt_sj}
\end{figure}

\noindent Theme: Mobility
\begin{enumerate}[a.]
    \item Overall, system VMT decreases with SOT and SOF compared to baseline UET.  
    \item VHD decreases with SOT and increases with SOF. SOF shifts traffic from highways to local roads in an attempt to save fuel as can be seen in Figure \ref{fig:vmt_sj}. In SOT, the total system delay reduction happens by rearranging flows from highways to throughways instead of neighborhood streets.
    
    \item Congested network miles in the morning peak decreases with SOT and increases with SOF consistent with delay increase.  
    \item Average trip length remains similar for all three cases. However, trip delay increases with SOF. 
\end{enumerate}

\noindent Theme: Equity
\begin{enumerate}[a.]
    \item Minority schools bear disproportionate impacts of traffic exposure. While impacting all types of schools, it is estimated to be more prevalent for minority schools. 22\% of schools are categorised as minority schools in San Jose. The proportion of schools affected by this predicted exposure is 41\% in baseline. This percentage further increases by at least 5 percentage points with SOT and SOF.
    \item Consistent with VMT reduction trends, VMT in disadvantaged communities reduces with SOT and SOF routing. Delay also shows similar trends. Note that the population in disadvantaged communities constitutes 32\% of the total, but account for 40\% VMT and 55\% VHD with UET.This reflects the tendency of these communities to be located near highways with high flow rates. These percentages are reduced with SOT and SOF optimizations.
\end{enumerate}

\begin{figure}[ht!]
    \centering
    \includegraphics[width=1\textwidth]{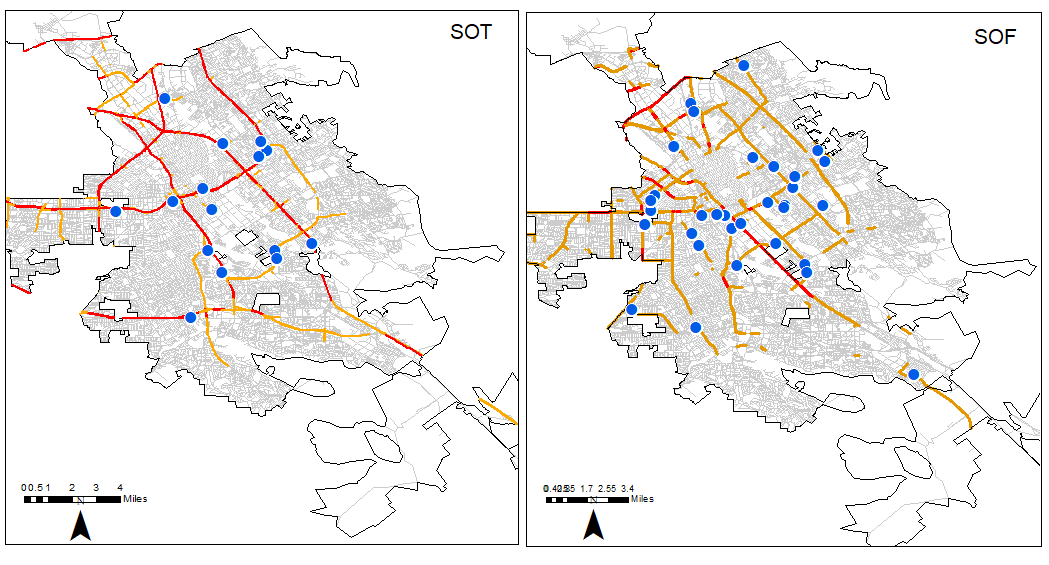}
    \caption{Theme Neighborhood: The figure illustrates the schools affected by high and medium traffic volume. Red links have an ADT greater than 50,000 and yellow links have ADT between 25,000 and 50,000.}
    \label{fig:schools_sj}
\end{figure}

\noindent Theme: Environment
\begin{enumerate}[a.]
    \item As expected, SOF produced  the lowest total system fuel consumption. SOT and baseline values are similar. Average trip fuel also shows similar trends with total system fuel consumption. These indicators are very sensitive to the fuel model used in the Mobiliti simulation, which currently considers only an average speed per link. We plan to improve this model in the future to account for speed variability that will be important for local link modeling.
\end{enumerate}

\newpage
\subsubsection{Comparison Across Cities}
\begin{figure}[ht!]
    \centering
    \includegraphics[width=.9\textwidth]{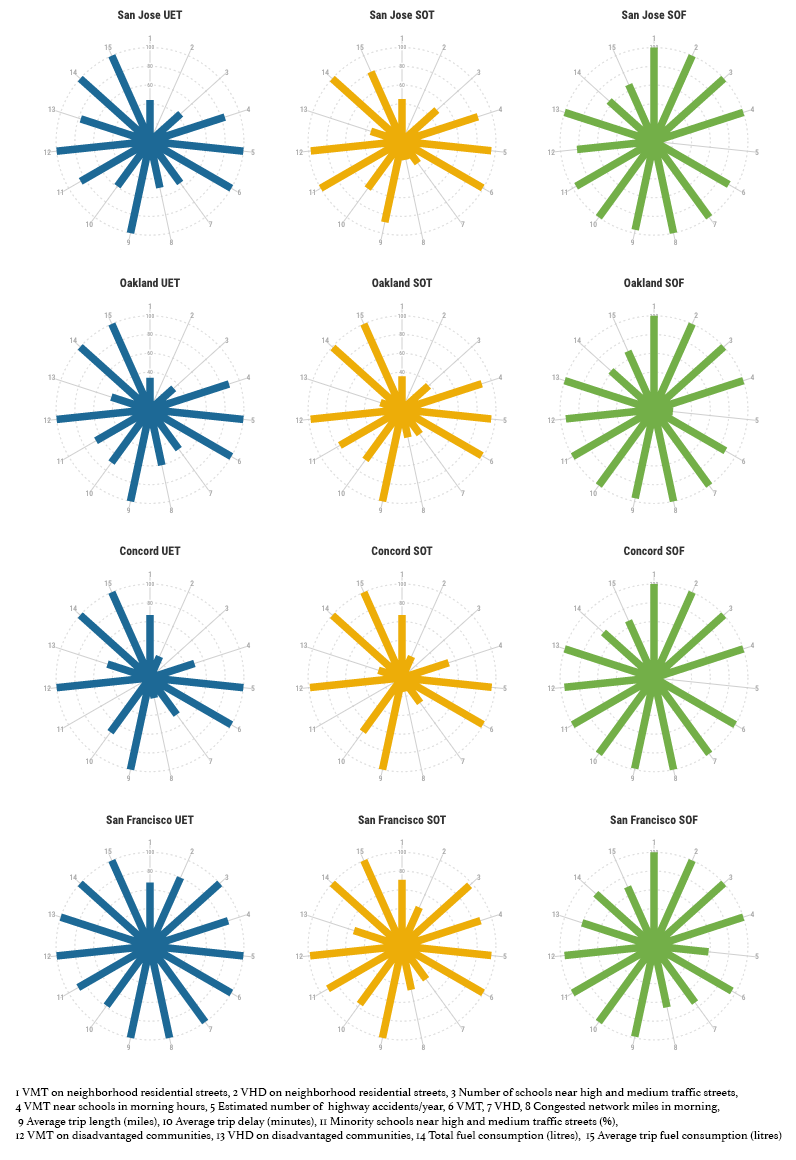}
    \caption{Comparison of SAEF metrics across cities for different route optimizations}
    \label{fig:allcities}
\end{figure}

Size, structure, land use, and density can vary widely across cities. By comparing routing optimizations across cities, similarities and differences can be seen with respect to our framework indicators which are listed below \ref{fig:allcities}.

\vspace{\baselineskip} 
\noindent Key Similarities:
\begin{enumerate}[a.]
    \item  For all four cities, percentage VMT carried by neighborhood residential streets increases with SOT and SOF compared to baseline. The highest increase is for Oakland, followed by San Jose.  
    
    \item For all cities but Concord, neighborhood residential streets VHD decreases with SOT and increases with SOF. However the percentage share out of the total system delay increases in both cases for all cities. With SOT the percentage increase is small and with SOF, the delay significantly increases for all cities with Oakland facing the worst impacts. Neighborhood streets in Oakland are predicted to experience 30 times higher delay as compared to baseline. Oakland has higher highway miles within the city limits and with SOF routing traffic to local roads, it does not have as much capacity on its local roads to absorb the traffic and thus would experience high neighborhood residential delays. For San Francisco and Concord the percentage delay increase is not as high as the other two cities. San Francisco has the lowest percentage of neighborhood residential streets in the network and its gridiron structure may account for the different flow dissipation.
    \item San Jose and Oakland are predicted to have a higher number of schools exposed to high and medium traffic with SOF compared to baseline. 
    \item The total system VMT decreases with SOT and SOF compared to baseline for all cities. 
    \item System level VHD decreases with SOT for all cities. VHD increases with SOF for all except San Francisco. 
    \item For the equity theme indicators - VMT and VHD on disadvantaged communities -  in San Jose, San Francisco and Concord, disadvantaged communities bear a disproportionate share of traffic compared to its population. For San Jose population in disadvantaged census tracts account for 32\% of the total population, but their streets carry 40\% of total VMT and 55\% total VHD with baseline. Similar trends can be observed in San Francisco and Concord. No significant change can be seen in the VMT proportion with SOT and SOF.
\end{enumerate}

\newpage
\newpage
\noindent Key Differences: 
\begin{enumerate}[a.]
    \item For Concord, the amount of neighborhood residential street VHD remains same for SOT and baseline, while it reduces for all other cities. Congestion was very low in the baseline UET case for Concord as such system level optimization had little effect. This is likely due to the spatial spread and low-density character of the city.
    \item Concord also differed in the number of schools exposed to high traffic - with no impacted schools in baseline and SOT and 1 school with SOF. For San Francisco, the number of affected schools remains similar for all the routing cases. Note that only 24\% schools in San Francisco are minority, but they account for 47\% of the schools exposed to high and medium traffic.

    \item COCs account for a higher percentage of the Oakland road network than the other three cities. As a consequence,a good percentage of the VMT and VHD occurs in these communities, 55\% and 50\% of total VMT and VHD respectively.
\end{enumerate}

\vspace{\baselineskip} 
Our intent with this work is to provide an evaluation framework to enable reflection on the consequence of policies, traffic management strategies and network changes. With an ability to model out proposed traffic management strategies, the planner can consider the trade-offs and potential unintended consequences. Realizing that there will always undesirable consequences of a specific strategy due the complexity and interconnectedness of transportation systems, the planner can then develop mediation strategies for the predicted results.

\section{Conclusion and Future work}
We have presented a framework to holistically evaluate the impacts of traffic management policies. We specifically focus on routing strategies using our key themes of neighborhood, safety, mobility, equity, and environment. A road network typology classification scheme that allows for the development of improved indicators is presented. Three traffic assignment optimization strategies are evaluated to offer insights into how the resulting routes and consequent traffic dynamics will impact the constituents of a city. Four chosen cities in the Bay Area were assessed for comparison purposes. We found that a) neighborhood residential streets are the most impacted with changes in routing strategies in terms of both increased traffic flow and delay, b) strategies differ in their traffic exposure impacts to schools, with minority schools bearing disproportionate impacts of traffic exposure, and c) the dynamics differ based on the network characteristics of a city e.g. percentage of highways, street layouts, density of roads. The framework creates an assessment system that will enable cities to evaluate the effects of transportation policies and management strategies.  It provides a tool for cities to assess their decisions in a comprehensive manner, while recognising the trade-offs resulting from the traffic dynamics. In future work, we plan to extend our analysis to include more cities and classify them based on their network characteristics to determine if patterns of traffic distribution/dissipation based on network characteristics exist. More importantly, we intend to use the framework to develop aggregate measures of safety and neighborhood that will drive novel socially-aware routing strategies and road network adjustments.  

\section*{Acknowledgment}
The work described here was sponsored by the U.S. Department of Energy (DOE) Vehicle Technologies Office (VTO) under the Big Data Solutions for Mobility Program, an initiative of the Energy Efficient Mobility Systems (EEMS) Program.  This research used resources of the National Energy Research
Scientific Computing Center, a DOE Office of Science User Facility
supported by the Office of Science of the U.S. Department of Energy
under Contract No. DE-AC02-05CH11231. 

\clearpage
\appendix
\setcounter{table}{0}
\renewcommand{\thetable}{\arabic{table}}
\subsection{SAEF Indicator Description }
\label{lab:appendixA}

\footnotesize\begin{longtable}{|p{2cm}|p{3cm}|p{5cm}|p{1.5cm}|p{2cm}|p{2cm}|}
    \caption{SAEF Indicator Description}\\
    \hline
    \textbf{Theme} & \textbf{Indicator} & \textbf{Description} & \textbf{Unit} & \textbf{Spatial level} & \textbf{Temporal level} \\

    Neighborhood & VMT on neighborhood residential streets & Vehicle miles travelled on neighbourhood residential street is calculated. City streets needs to be classified based on complete streets guidelines to identify neighborhood residential streets based on land use and transport functionalities & miles & Community & Entire day / Peak hours \\
    
    Neighborhood & VHD on neighborhood residential streets & Vehicle hours of delay on neighbourhood residential street is evaluated. City streets needs to be classified based on complete streets guidelines to identify neighborhood residential streets & hours & Community & Entire day / Peak hours \\
    
     Neighborhood & Number of schools near high and traffic streets & High traffic is considered as streets with average daily traffic (ADT) greater than 50,000 and medium traffic streets has ADT greater than 25,000 vehicles per day. All roads within 250 m of school vicinity is considered for this analysis as traffic on them has the most impact on schools. & number & City & Entire day \\
     
     Neighborhood & VMT near schools in the morning hours & Vehicle miles travelled in the 250 meter vicinity of schools.  VMT is calculated for 7-8am in the morning. This indicator helps in understanding the difference in traffic flow around schools for various routing strategies & miles & City & 	Peak hours\\
     
    Neighborhood & Average travel speed along neighborhood residential streets	& Average travel speed on neighborhood residential streets can be calculated for morning or evening peaks. This gives an estimate of how the traffic moves in neighborhoods  & miles per hour (mph) & Community & Entire day / Peak hours \\
    
    Neighborhood & Percentage vehicles above speed limit in neighborhood streets & Percentage vehicles travelling above the posted speed limit at different times of day gives an indication of the traffic flow affecting the quality of life in neighborhoods. & percentage (\%) & Community & Peak hours\\
    
    Safety & Estimated number of  highway accidents/year & Highway accidents estimated based on traffic volume and road geometry. Safety performance functions for highways can be used. & number &	City & 	Entire day \\
    
    Safety & Estimated urban accidents/year & Urban accidents estimated from traffic volume and road geometry. & number	& City &  	Entire day \\
    
    Safety & VMT on high pedestrian and ATM streets &	Vehicle miles travelled on active transport mode users and pedestrians.  This measures the multi modal safety of roads. Routing strategies that increases VMT on these roads hinders others active users of the road. &	miles	& City &	Entire day / Peak hours\\
    
    Safety & VMT on high risk streets	& High risk roads involves roads with hot spots, steep slopes etc which can be dangerous when unknown drivers are routed through them. &	miles&	City &	Entire day / Peak hours \\
    
    Mobility & VMT&	Total system vehicle miles travelled &	miles&	City &	Entire day \\
    
    Mobility & VHD &	Total system vehicle hours of delay&	hours&	City &	Entire day \\
    Mobility & Congested network miles in morning&	Roads with volume over capacity greater than or equal to 1 during morning hours. This measures the congested network for each routing strategy.  Spatial spread of congestion might be good to visualise as well. &	miles&	City &	Peak hours\\ 
    Mobility & Average trip length &	Average trip length for a city. It can segregated to work and non work trips as well.& 	miles &	Individual & Entire day\\
    Mobility & Average trip delay&	Average trip delay for a city. It can segregated to work and non work trips as well.& minutes	& Individual / City & Entire day  \\
    Mobility & Bicycle/Pedestrian LTS	& Bicycle and pedestrian level of traffic stress rating can be given to road segments based on the type of facility, traffic volume and speed. Standard methods from literature can be adopted. &		&  City / Community &	Peak hours\\
    
    Equity & Minority schools near high and medium traffic streets &	Minority schools are identified as majority children in school in minority classification. Then these schools are checked for a buffer of 250 metres to identify exposure to high and medium traffic. &	number or percentage (\%) &	City &	Entire day \\
    Equity & VMT on disadvantaged communities&	Vehicle miles travelled on disadvantaged communities. Disadvantaged communities can be identified based on each city's or county's definition. &	miles& Community &	Entire day \\
    Equity & VHD on disadvantaged communities & Vehicle hours of delay on disadvantaged communities. Disadvantaged communities can be identified based on each city's or county's definition. &hours & Community &	Entire day \\
    Equity & \% people within 45 min of one way work commute by demography & Percentage population within 45 minutes of work commute helps in understanding the accessibility of a place. Based on which routing strategies objective and trade offs can be explored.& percentage (\%) & 	City &	Entire day\\
    
    Environment & Total fuel consumption & Total system fuel consumption by total trips. & litres&	City &	Entire day \\
    Environment & Average fuel consumption/trip	& Per trip average fuel consumption. & litres & Individual &	Entire day \\
    Environment &Emissions percapita & Emissions per person in a city. Methods to calculate emissions from vehicular traffic can be adopted from the literature. &  & City  & Entire day \\
    \hline
    \label{tab:saef_detailed}
\end{longtable}

\subsection{QDTA Formulations}
\label{lab:appendixB}

\begin{equation}
\centering
    UET: \sum_{a \in A} \int_0^{f_a} c_a(s) \ ds
\label{eq:cost_function1}
\end{equation}

\begin{equation}
\centering
    SOT: \sum_{a \in A} f_a c_a(f_a)
\label{eq:cost_function2}
\end{equation}

\begin{equation}
\centering
    SOF: \sum_{a \in A} f_a m_a(v_a)
\label{eq:cost_function3}
\end{equation}

where:

$c_{a}(f_a) = c_{0,a}(1+\alpha(\tfrac{f_a}{C_a})^{\beta})$
\\
$c_a(f_a)$ is the travel time on link $a$; $f_a$ is the traffic flow assigned to link $a$; $c_{0,a}$ and $C_a$ are the free-flow travel time and capacity associated with the link;  $\alpha$ and $\beta$ selected are BPR parameters 0.15 and 4 respectively.\\

\noindent $m_a(v_a) = l_a ( A + \frac{B} {v_a}+ C v_a^2)$\\
\(m_a(v_a)\) is the fuel consumption on link \(a\); $l_a$ is the length of link \(a\); $v_a$ is the link traversal speed calculated from BPR; $A,B,C$ are parameters estimated from Argonne National Laboratory drive cycle data $A = -0.00654170, B = 1.902150, C = 0.00001588.$ 

\subsection{Bay Area Network}
\label{lab:appendixC}
Bay Area network functional classification is provided in Table \ref{tab:fcclass} and the network typologies are provided in Table \ref{tab:link_class}

\begin{table}[h]
\centering
\footnotesize
\caption{Functional Road Classes}
\label{tab:fcclass}     
\begin{tabular}{|p{2cm}|p{12cm}|}
\hline
\noalign{\smallskip}
Functional Class & Definition \\
\noalign{\smallskip}
\hline
\noalign{\smallskip}
1 & Allowing for high volume, maximum speed traffic movement\\
2 & Allowing for high volume, high speed traffic movement \\
3 & Providing a high volume of traffic movement \\
4 & Providing for a high volume of traffic movement at moderate speeds between neighbourhoods \\
5 & Roads whose volume and traffic movement are below the level of any other functional class \\
\noalign{\smallskip}
\hline
\end{tabular}
\end{table} 

\begin{table}[ht!]
\centering
\footnotesize
\caption{Network Typologies for Bay Area}
\label{tab:link_class}       
\begin{tabular}{|p{1cm}|p{5cm}|p{3cm}|p{5cm}|}
\hline
\noalign{\smallskip}
Sl No & Street Type & Length (thousand miles) & Remarks \\
\noalign{\smallskip}
\hline
\noalign{\smallskip}
1 & Neighborhood Residential Street & 26 & Neighborhood streets with adjoining residential land use\\
2 & Residential Throughway & 2.9 & Throughway streets with adjoining residential land use \\ 
3 & Neighborhood Commercial Street & 1.8 & Neighborhood streets with adjoining commercial land use \\
4 & Commercial Throughway & 1.1 & Throughway streets with adjoining commercial land use \\
5 & Industrial Street & 1.7 & Neighbourhood or Throughway streets with adjoining industrial land use \\
6 & PSP Street & 2.5 & Neighbourhood or Throughway streets with adjoining public and semi public land use \\
7 & Highway & 3.4  & Highways including ramps \\
8 & Others & 9.9 & Includes streets with adjoining land use as green spaces, undeveloped land,parking lots, water bodies etc.   \\
\noalign{\smallskip}
\hline
\end{tabular}
\end{table}

\subsection{SPF Parameters Used for California Highways}
\label{lab:appendixD}
The safety performance function parameters estimated from PEMS data for the state of California is provided in Table \ref{tab:spf_param} \cite{shankar_methods_2015}.
\begin{table}[ht!]
\footnotesize
\centering
\caption{SPF Parameters for California Highways}
\label{tab:spf_param} 
\begin{tabular}{|p{3cm}|p{3cm}|p{3cm}|}
\hline
\noalign{\smallskip}
Number of lanes & alpha	& beta\\
\noalign{\smallskip}
\hline
\noalign{\smallskip}
1 &	-7.09 &	0.98\\
2 &	-7.09 &	0.98\\
3 &	-7.09 &	0.98\\
4 &	-5.78 &	0.82\\
5 &	-6.49 &	0.89\\
6 &	-6.49 &	0.89\\
7 &	-6.49 &	0.89\\
8 &	-10.75 & 1.24  \\
\noalign{\smallskip}
\hline
\end{tabular}
\end{table}

\subsection{Summary of Results }
\label{lab:appendixE}
\begin{figure}[ht]
    \centering
    \includegraphics[width=0.9\textwidth]{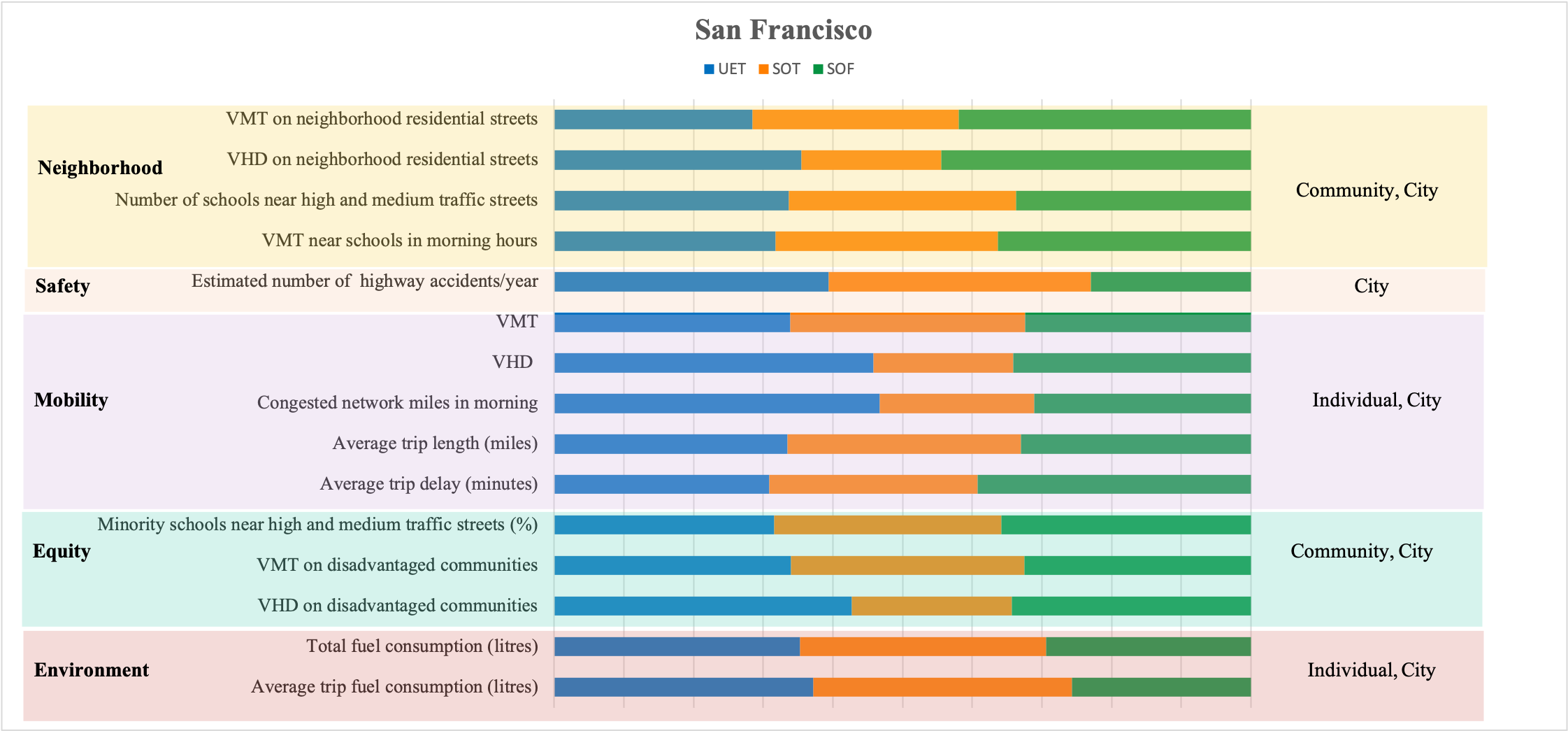}
    \includegraphics[width=0.9\textwidth]{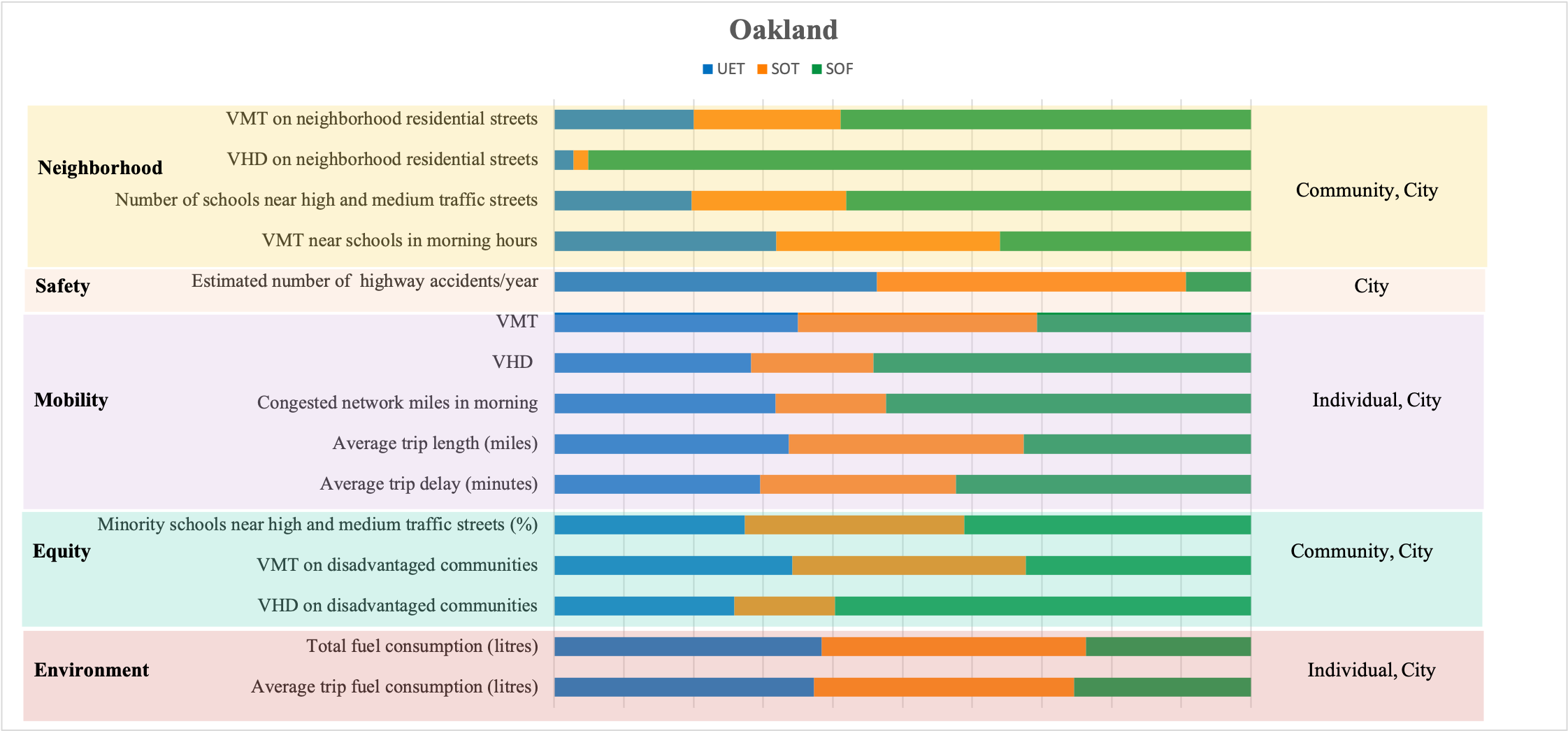}
    \includegraphics[width=0.9\textwidth]{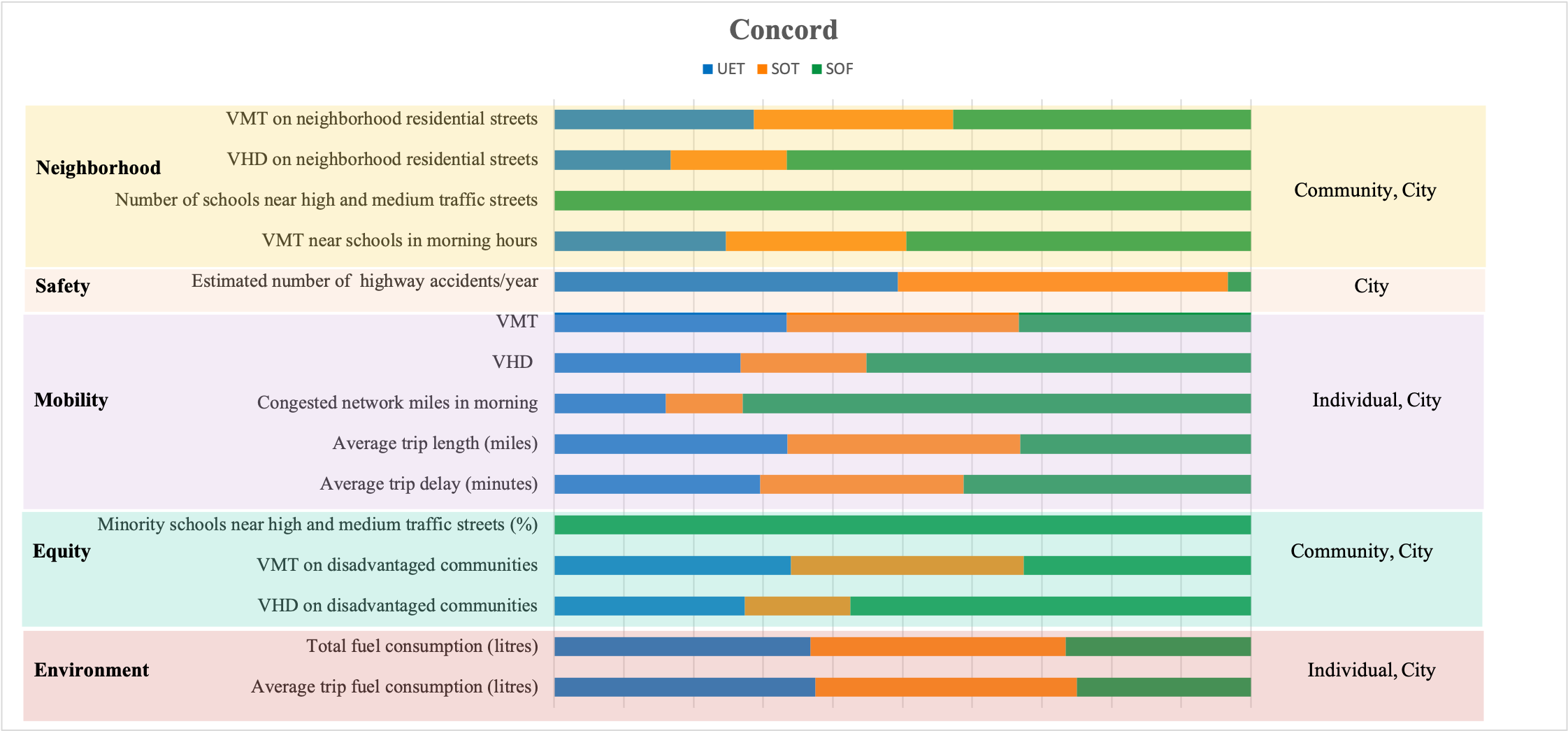}
    \caption{SAEF results for San Francisco, Oakland and Concord}
    \label{fig:SAEF_three}
\end{figure}

Summary of results is provided in Table \ref{tab:systemmetricsfc} for the case study cities. Figure\ref{fig:SAEF_three} shows the SAEF charts for the three cities.  

\begin{sidewaystable}
\centering
\scriptsize
\caption{Summary of Results: SAEF for case study cities}
\label{tab:systemmetricsfc} 
  \begin{tabular}{|p{0.5cm}|p{4cm}|p{1cm}|p{1cm}|p{1cm}|p{1cm}|p{1cm}|p{1cm}|p{1cm}|p{1cm}|p{1cm}|p{1cm}|p{1cm}|p{1cm}|} 
    \hline
    \multirow{2}{*}{\textbf{Sl}} & \multirow{2}{*}{\textbf{Indicator}} & \multicolumn{3}{|c|}{\textbf{San Jose}} & \multicolumn{3}{|c|}{\textbf{San Francisco}} & 
    \multicolumn{3}{|c|}{\textbf{Oakland}} & \multicolumn{3}{|c|}{\textbf{Concord}} \\
    \cline{3-5} \cline{6-8} \cline{9-11}\cline{12-14}
  & & UET  & SOT & SOF & UET  & SOT & SOF & UET  & SOT & SOF & UET  & SOT & SOF \\ 
    \hline
    1 & VMT on neighborhood residential streets (in thousands) & 846 &	 867 & 	 1,922 &	 471 &	 493&	 1,383 &	 96& 	 96 &	 144 &	 1,104 &	 1,152 &	 1,628 \\
    2 & VHD on neighborhood residential streets & 62 &	 41 &	 900 &	 32& 	 24 	& 1,085 &	 1 &	 1 &	 4& 	 620& 	 351& 	 776 \\
    3 & Number of schools near high and medium traffic streets & 22& 	 25& 	 50& 	 17& 	 19 &	 50 &	 0 &  	 0 &  	 1 &	 32 	& 31 &	 32 \\
    4 & VMT near schools in morning hours & 109,467 &	 111,090 &	 130,674& 	 112,712& 	 113,790& 	 127,388 &	 3,263 &	 3,430 &	 6,557 &	 110,035 &	 110,617 &	 125,779 \\
    5 & Estimated number of  highway accidents/year & 4,793 &	 4,582 	& 170 &	 3,002 	& 2,869 	& 608 &	 537 &	 516 &	 36 &	 1,519 &	 1,453 	& 887 \\
    6 & VMT (in thousands)  &18,883 &	 18,741 &	 17,374 &	 9,143&	 8,978 &	 8,028 &	 2,195 &	 2,186 &	 2,190	& 8,994& 	 8,966& 	 8,607\\
    7 & VHD & 10,770 &	 5,639& 	 19,651 &	 7,617& 	 4,714 	& 14,616& 	 565 &	 384 &	 1,167 	& 17,355 	& 7,584 	& 12,917 \\
    8 & Congested network miles in morning & 43 	& 17 &	 85 &	 20 &	 10 	& 33 &	 1& 	 1 	& 5 &	 21 &	 10 &	 14 \\
    9 & Average trip length (miles) & 8 &	 7 &	 8 	& 9 &	 9 &	 9& 	 9 &	 9& 	 9 &	 7& 	 7 &	 7 \\
    10 & Average trip delay (minutes) & 10 &	 11 &	 17 &	 12 &	 11& 	 17 &	 11& 	 11 &	 16 &	 11 &	 11 &	 14 \\
    11 & Minority schools near high and medium traffic streets (\%) & 41 &	 48 	& 46& 	 41 &	 47 &	 62 	& 0&  	 0&  	 100& 	 47 &	 48& 	 53 \\
    12 & VMT on disadvantaged communities (in thousands) & 7,279 &	 7,139	& 6,006& 	 5,041& 	 4,950& 	 4,769& 	 401& 	 396 &	 385& 	 2,704 &	 2,681	& 2,592 \\
    13 & VHD on disadvantaged communities & 5,811 	& 2,684 	& 7,585& 	 3,797 &	 2,222 &	 9,000 &	 128 &	 70 &	 267 	& 3,335& 	 1,877 &	 2,744 \\
    14 & Total fuel consumption (litres)&  3,520,859 &	 3,519,303& 	 2,267,914 &	 1,700,216 &	 1,683,629 	& 1,049,879 	 &405,936& 	 404,637 	& 294,039& 	 1,295,692 &	 1,294,413 &	 1,081,003 \\
    15 & Average trip fuel consumption (litres) & 2 &	 1 &	 1& 	 2& 	 2 &	 1& 	 2 &	 2& 	 1& 	 1 &	 1 &	 1 \\
    \hline
  \end{tabular}
\end{sidewaystable}

\clearpage
\nocite{*}
\bibliographystyle{IEEEtran}
\bibliography{IEEEabrv,references}

\end{document}